\documentclass[prl,twocolumn,floatfix]{revtex4}
\usepackage{graphicx}
\usepackage{graphics}
\usepackage{latexsym}
\usepackage{amsmath, amsthm, amssymb, wasysym}
\usepackage{xcolor}
\usepackage{epsfig}
\usepackage{hyperref}
\usepackage[normalem]{ulem}

\begin{document}
\title{Universal to Non-Universal Transition of the statistics of
Rare Events During the Spread of Random Walks}

\author{R. K. Singh}
\email{rksinghmp@gmail.com }
\affiliation{Department of Physics, Bar-Ilan University, Ramat-Gan 5290002, Israel}

\author{Stanislav Burov}
\email{stasbur@gmail.com }
\affiliation{Department of Physics, Bar-Ilan University, Ramat-Gan 5290002, Israel}

\begin{abstract}
Particle hopping is a common feature in heterogeneous media.
We explore such motion by using the widely applicable formalism of the continuous time
random walk and focus on the statistics of rare events. 
Numerous experiments
have shown that the decay of the positional probability density function $P(X,t)$, describing the statistics of rare events, exhibits universal exponential decay. We
show that such universality ceases to exist once the threshold of
exponential distribution of particle hops is crossed. While the mean
hop is not diverging and can attain a finite value; the transition itself is critical. The exponential universality of rare events arises due to the contribution of all the different states occupied during the process. Once the reported threshold is crossed, a single large event determines the statistics.  In this realm, the big jump principle replaces the large deviation principle, and
the spatial part of the decay is unaffected by the temporal properties of rare events.
\end{abstract}

\maketitle

\newcommand{\fr}{\frac}
\newcommand{\tl}{\tilde}
\newcommand{\avg}{\langle N_t \rangle}
\newcommand{\mean}{\langle \tilde{N}_s \rangle}

According to Wikipedia, "Rare or extreme events are events that occur with low frequency,
and often refers to infrequent events that have widespread impact and might destabilize systems".
Notable examples of rare events include the stock
market crash \cite{amihud1990liquidity}, earthquakes \cite{lomnitz1966statistical} and cyclones
\cite{leckebusch2004relationship}. 
The most recent on the list is the
2020 pandemic, which brought the whole world to a standstill.
The frequency and the scale of rare events in each field are very important. If the
rare events are not "too rare" and large enough, then the usual statistical behavior is
completely dictated by such rare events. A perfect example of situations when rare events
fundamentally modify the nature of a physical process is sub-diffusion \cite{metzler2000random,
weiss1983random,alexander1981excitation,havlin1987diffusion,isichenko1992percolation,
shlesinger1993strange,haus1987diffusion}.
While for normal diffusion, the mean squared displacement (MSD) of a tracked particle grows
linearly with time, for sub-diffusion, the MSD grows in a non-linear fashion, i.e.
$\langle X^2 \rangle \sim t^{\alpha}$, while $0<\alpha<1$.
Sub-diffusion generally occurs when there are long time or space correlations present. 
One of the sources for such correlations can appear in the form of an extremely long sojourn
times, when the tracked particle is trapped in a specific region in space \cite{scher1991time}.
When sub-diffusion
occurs in amorphous materials like glasses \cite{binder1986spin}
or living cells \cite{manzo2015weak,graneli2006long,wang2006single}, it is quite
common to observe trajectories that are dominated by several (and sometimes even one)
very long trapping events \cite{jeon2011vivo}. 

In many cases, the appearance of sub-diffusion is accompanied by unusual physical phenomena:
aging \cite{bouchaud1992weak,akimoto2016universal,barkai2003aging},
weak ergodicity breaking \cite{bouchaud1992weak,rebenshtok2007distribution,bel2005weak}
and non-self-averaging \cite{bouchaud1992weak,burov2007occupation,pronin2022non,sabhapandit2011record,burov2017quenched,shafir2022case}.
Studies of theoretical models where trapping is present, like the continuous time random walk
(CTRW) \cite{klafter2011first} and the quenched trap model (QTM) \cite{bouchaud1990anomalous}
show that there is a critical transition to sub-diffusion, aging, and non-ergodic behavior,
when the mean trapping time is diverging.
This transition is also accompanied by
a transformation of the positional probability density function (PDF).
The universal Gaussian center of the positional PDF turns into
an $\alpha$-stable L\'{e}vy type \cite{uchaikin2003self,garoni2002levy}, i.e. shape that depends on a specific parameter $\alpha$ that determines the trapping times.
This transition between
a universal PDF that is determined by an accumulation of many events, and a non-universal
PDF that follows the properties of one single rare event is a key feature associated with the
the appearance of new physical phenomena. 
Can transitions from universal to non-universal PDFs symbolize an underlying phase transition that might lead to uncovered surprising physics, like in the case of sub-diffusion? What are the properties of single rare events that can cause such a critical transition in behavior? 
In this work, we focus on the appearance of the universal to the non-universal transition of
a positional PDF that, unlike the case of sub-diffusion, appears for large positions $|X|$ (the tails of the PDF).



Recently, in a huge number of experiments, exponential (and not Gaussian) decay of the PDF for large $|X|$
was spotted.
Such decay of the PDF is termed Laplace tails \cite{chechkin2017brownian}.
Notable examples include colloidal beads in heterogeneous optical force
field \cite{pastore2021rapid}, zooplanktons under crowding \cite{uttieri2021homeostatic},
glass-forming liquids \cite{rusciano2022fickian}, nanoparticles in polymer melts
\cite{xue2016probing}, motion of particles at liquid-solid interface \cite{wang2017three},
particle displacements close to glass and jamming transitions \cite{chaudhuri2007universal,chaudhuri2008random}
etc. 
In \cite{wang2012brownian} it was suggested that there might
exist a universal convergence to the Laplace tails. 
Indeed it was proven, by exploiting the CTRW model that for a wide
range of processes, the tails of the PDF are determined by the accumulation of many events and follow exponential
decay (up to logarithmic corrections) \cite{barkai2020packets,wang2020large,pacheco2021large}. 
So by following the presented discussion, we ask:
Is there a transition from universal Laplace tails to non-universal and process-specific
tails? And if so, is it accompanied by a critical transition like the diffusion to subdiffusion
transition? Are the tails of a PDF determined by one occurrence of one single event
\cite{foss2013heavy,embrechts2013modelling,kutner2002extreme,
de2013asymmetric,majumdar2005nature}
or they represent an accumulation of many realizations of not-so-large but more frequent events
\cite{hollander2000large,ellis1999theory,dembo2009large,nickelsen2018anomalous,touchette2009large}? 

\begin{figure}
\includegraphics[width=0.5\textwidth]{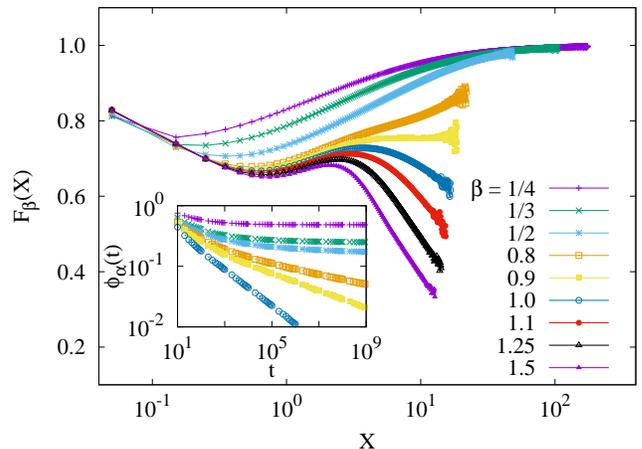}
\caption{$F_\beta(X)$ for a CTRW with jumps $x \sim f(x)$ with mean zero and
variance one as a function of $\beta$. The waiting time distribution
$\psi(\tau) = e^{-\tau}$ and the measurement
is done at $t = 0.6$ for trajectories reaching a position $\pm X$. Inset shows the
behavior of $\phi_\alpha(t)$ for $\psi(\tau) \sim \tau^{-1-\alpha}$ for large $\tau$
and $\alpha = 0.5, 0.8, 0.9, 1.1, 1.2, 1.5$ from top to bottom.}
\label{fig1}
\end{figure}

Similar to the case of subdiffusion to diffusion transition, we use the celebrated CTRW model,
originally exploited for explaining motion in amorphous materials 
\cite{binder1986spin,manzo2015weak,graneli2006long,wang2006single} and exponential decay of tails \cite{chechkin2017brownian}.
In this model, a
particle performs random jumps in space and waits for
a random amount of time between every two jumps. All the jumps and waiting times are independent and identically distributed (IID)
random variables. The distribution of a jump $x$ is given by $f(x)$, while each waiting
time $\tau$ has a distribution $\psi(\tau)$. The position of the process $X$ at time $t$
is determined by the random number of jumps $N_t$ performed by time $t$, i.e. $X=x_1+\dots+x_{N_t}$ where $x_i$, is the size of a single jump.
The positional PDF $P(X,t)$ is readily obtained in terms of the subordination equation
by conditioning on the number of jumps $N$,
\cite{he2008random,magdziarz2008equivalence,bouchaud1990anomalous,metzler2000random}
\begin{align}
	P(X,t) = \sum^\infty_{N=0}P_N(X)Q_t(N)
	\label{subor}
\end{align}
where $P_N(X)$ is the distribution of $X$ for a given $N$($=N_t$ for a fixed measurement time $t$) and $Q_t(N)$ is the distribution of
the number of jumps up to time $t$.
The mentioned phenomena, like anomalous transport and aging, appear for CTRWs with
$\psi(\tau) \sim \tau^{-1-\alpha}$ ($\tau\to\infty$) when $\alpha < 1$.
For such $\psi(\tau)$,  the sum $\tau_1+\cdots+\tau_{N_t}$ is dominated (in the $t \to \infty$ limit)
by the maximal summand \cite{derrida1997random},
as can be observed from the behavior of
\begin{align}
\phi_\alpha(t) = \Big\langle \fr{\text{max}\{\tau_1, ..., \tau_{N_t}\}}{\sum^{N_t}_{i=1} \tau_i}\Big\rangle
\label{phi_t}
\end{align}
for large $t$.
While the maxima of a set of random variables has been extensively
studied \cite{majumdar2010universal,mounaix2018asymptotics,mori2020distribution,holl2021big},
the definition of $\phi_\alpha(t)$ has the advantage that it ranges from infinitesimal
to finite values. This makes it appropriate as an order parameter.
We see in
Fig.~\ref{fig1} (inset) that $\phi_\alpha(t)$ saturates to a finite value
for $\alpha = 0.5,~0.8,~0.9$ while for $\alpha = 1.1,~1.2,~1.5$ exhibits a decaying behavior
at large times. This difference in the properties of $\phi_\alpha(t)$ follows from the fact
that for $\alpha < 1,~ \langle \tau \rangle = \int^\infty_0 d\tau ~\tau \psi(\tau) \to
\infty$, while it is finite for $\alpha > 1$.
This implies that as we move from $\alpha > 1$ to $\alpha < 1$,
the rare fluctuations in the sequence of waiting times $\{\tau_1, ..., \tau_{N_t}\}$
exhibit qualitatively different behaviors. Consequently, the thermodynamic limit in
which the system behaves extensively (all waiting times $\tau_i$ are of the same order
of magnitude) ceases to exist for $\alpha < 1$. Can we see similar
behavior if we focus on spatial fluctuations? This question comes up naturally once
we realize that the system has access to the entire phase
space in the thermodynamic limit. The phase space can be swept by focusing on large spatial
fluctuations without going to the $t \to \infty$ limit.
This means we take the large $|X|$ limit and expect the system to visit many states during its evolution.



Motivated by this, let us look at a CTRW described by Eq.~\eqref{subor} with jumps following
$f(x) \sim e^{-|x|^\beta}$ with $\beta > 0$ \cite{nardon2009simulation}. In analogy
with $\phi_\alpha(t)$ let us define
\begin{align}
        F_\beta(X) = \Big\langle\fr{\text{max}\{|x_1|,...,|x_{N_t}|\}}
        {\sum^{N_t}_{i=1}|x_i|}\Big\rangle.
\end{align}
We see in Fig.~\ref{fig1} that even for a well behaved $\psi(\tau)$ like the exponential
distribution, $\lim_{|X| \to \infty}F_\beta(X)$ decreases monotonically for $\beta \ge 1$ and
takes a finite value for trajectories with $\beta < 1$. In other words,
\begin{align}
        \lim_{|X| \to \infty}F_\beta(X) =\begin{cases}
                0,~~\beta \ge 1 \\ 
                c_\beta,~~\beta < 1
        \end{cases}
\end{align}
where $c_\beta$ is a nonzero constant that depends on $\beta$. Although $c_\beta$ is seen
to approach unity for $\beta = 1/2$ and lower values, $\beta = 0.9$ seems to saturate
to a value less than one. 
While the transition of $\phi_\alpha(t)$ is driven by the divergence of the mean waiting time $\langle \tau \rangle$, this is not the case for the transition of $F_\beta(X)$. 
All the moments $\langle x^n\rangle$ are finite for $\beta<1$. Nevertheless, the effective behavior of $\lim_ {t\to\infty}\phi_\alpha(t)$ and $\lim_{|X|\to\infty}F_\beta(X)$ is similar. As a function of the parameters $\alpha$/$\beta$, there is a transition from a state defined by the accumulation of many events to a state dominated by a single large event.



At this point, it is worth noticing that the role played by $\phi_\alpha(t)$ for
describing temporal fluctuations is taken over by $F_\beta(X)$ for spatial fluctuations.
While the emergence of a non-Gaussian center accompanying the transition of $\phi_\alpha(t)$
is well understood \cite{uchaikin2003self,garoni2002levy} the corresponding
behavior of $P(X,t)$ accompanying the transition of $F_\beta(X)$ is
not known. It was shown in Ref.~\cite{barkai2020packets} that
$P(X,t)$ exhibits universal tails exhibiting exponential decay for $\beta > 1$
and $\psi(\tau)$ analytic near $\tau=0$. Furthermore, just like $\phi_\alpha(t)
\to 0$ marks the emergence of a universal Gaussian center following the central
limit theorem, $F_\beta(X) \to 0$ at $|X| \to \infty$ characterizes the universal
exponential tails of $P(X,t)$ \cite{barkai2020packets}. On the other hand, a
finite value of $\phi_\alpha(t)$ for $\alpha < 1$ is reminiscent of the L\'{e}vy stable
PDF and is a reminder of the dynamical phase transition (DPT) \cite{nyawo2017minimal,nyawo2018dynamical,
garrahan2007dynamical,jack2013large} characterized solely by the rare fluctuations,
but in the temporal domain. Does it mean we can see a similar DPT at finite times
if we focus on the tails of $P(X,t)$? We  now answer this question by exploring the case of $\beta \le 1$.

For $\beta = 1$, the jumps follow Laplace distribution,
that is, $f(x) = \fr{a}{2}\exp(-a|x|)$.
This implies that the characteristic
 function of a single jump, defined as the Fourier transform
$\hat{\lambda}(k) = \int^\infty_{-\infty}dx ~e^{ikx}f(x) = \fr{a^2}{a^2+k^2}$ \cite{klafter2011first}.
Unlike the case for $\beta > 1$, $\hat{\lambda}(k)$ for $\beta = 1$ has poles
in the complex plane. Furthermore, since the jumps are IID, the distribution of a sum of $N$ jumps in Fourier space is $\hat{\lambda}^N(k)$ and inverting it via contour integration
\cite{watson1922treatise} we find that the
distribution of position $X$ after $N$ jumps is \cite{gradshteyn2014table}
\begin{align}
\label{PNX}
P_N(X) &= \fr{ae^{-a|X|}}{2^{2N-1}\Gamma(N)}\sum^{N-1}_{m=0}
\fr{(2N-m-2)!}{m!(N-m-1)!}(2a|X|)^m\nonumber\\
&\approx \fr{ae^{-a|X|}}{2^{2N}\Gamma(N)}\int^N_0 dm~\exp[K(N,m)]
\end{align}
where $K(N,m) = (2N-m)\log(2N-m) - m\log(m) - (N-m)\log(N-m) - (N-m) + m\log(2a|X|)$.
The last line follows from Stirling's approximation
$N! \approx N^N e^{-N}$. In order to evaluate the integral in (\ref{PNX}) we use
Laplace's method and locate $m_0$ such that $\fr{dK}{dm}|_{m=m_0} = 0$.
This implies $m_0 = a|X| + N - \sqrt{a^2X^2 + N^2}$.
In the limit $a|X|/N \gg 1$, 
$m_0 \approx N - \fr{N^2}{2a|X|}$ and then from (\ref{PNX}) the large
deviation form is obtained
\begin{align}
\label{ldf_pnx}
	P_N(X) \sim \exp\Big[-NI_N\Big(\fr{|X|}{N}\Big)\Big],
\end{align}
with the rate function $I_N\Big(\fr{|X|}{N}\Big) = -\fr{a|X|}{N}+\log\Big(\fr{a|X|}{N}\Big)$.
This implies that for large deviations, the distribution of the sum possesses exponentially
decaying tails with logarithmic corrections. For $\psi(\tau)$ analytic near zero
\begin{align}
\label{wt_time}
\psi(\tau) \stackrel{\tau \to 0}{\sim} C_A \tau^A + C_{A+1} \tau^{A+1} +
C_{A+2} \tau^{A+2} + \cdots,
\end{align}
where $A$ is a non-negative integer, $Q_t(N)$ admits a large deviation form
$Q_t(N) \stackrel{N \to \infty}{\sim} \exp[-NI_N(t)]$ with a universal rate
function \cite{burov2020limit} $I_N(t) = -\fr{C_{A+1}}{C_A}\fr{t}{N} - (A+1)\Big[1 + \log\Big
\{\fr{(C_A\Gamma(A+1))^{\fr{1}{A+1}}}{A+1}\fr{t}{N}\Big\}\Big]$.
Using $P_N(X)$ from (\ref{ldf_pnx}) and $Q_t(N)$ in (\ref{subor}) we have for large $N$
\cite{daniels1954saddlepoint}
\begin{align}
\label{pxt}
P(X,t) \approx \int^\infty_0 dN \exp[\kappa(N)] \approx
\sqrt{\fr{2\pi}{|\kappa''(N_0)|}}\exp[\kappa(N_0)],
\end{align}
where
\begin{align}
\label{KN}
\kappa(N) &= -a|X| + N\log\Big(\fr{a|X|}{N}\Big)\nonumber\\
&- Ct + N(A+1)\Big[1 + \log\Big(\fr{d_2}{N}\Big)\Big],
\end{align}
with $C = -\fr{C_{A+1}}{C_A}$ and $d_2 = \fr{[C_A\Gamma(A+1)]^{\fr{1}{A+1}}}{A+1}t$
and $N_0$ is the solution of $\kappa'(N_0) = 0$. For large $a|X|/N$ we have
\begin{align}
\label{fN0x}
	N_0 \approx \mu(a|X|)^{\fr{1}{A+2}}
\end{align}
with $\mu = d_2^\fr{A+1}{A+2}$. Using $\kappa''(N_0) = -\fr{A+2}{N_0}$ in (\ref{pxt}) we find
\begin{align}
\label{pxt_ldf}
	P(X,t) &\underset{|X|/t \to \infty}{\sim}
	\sqrt{\fr{2\pi}{A+2}\mu (a|X|)^{\fr{1}{A+2}}}\nonumber\\
	&\exp\Big[-t\Big\{C + \fr{a|X|}{t} - \Big(C_A\Gamma(A+2)\fr{a|X|}{t}\Big)^{\fr{1}{A+2}}\Big\}\Big]
\end{align}
\begin{figure}
\includegraphics[width=0.45\textwidth]{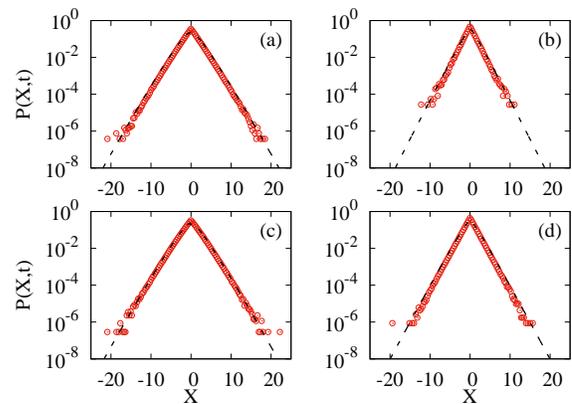}
\caption{Comparison of numerically estimated $P(X,t)$ (red circles) against the solution given
in (\ref{pxt_ldf}) (black dashed line). The waiting time distributions are the following:
(a) exponential mixture $\psi(\tau) = p_1 r_1 e^{-r_1 \tau} + p_2 r_2 e^{-r_2 \tau}$ with $r_1 = 1/4,
r_2 = 5/2, p_1 = 1/4, p_2 = 3/4$ at $t = 0.7$; (b) gamma distribution $\psi(\tau)
= \tau^3 e^{-\tau}/6$ at $t = 0.8$;
(c) half Gaussian distribution $\psi(\tau) = \sqrt{2/\pi}e^{-\tau^2/2}$ at $t = 1.5$;
(d) power law distribution $\psi(\tau) = 1/(1+\tau)^2$ at $t = 0.4$.}
\label{fig2}
\end{figure}\\
From Fig.~\ref{fig2} we see that the large deviation form of $P(X,t)$ evaluated in
(\ref{pxt_ldf}) agrees with numerically estimated $P(X,t)$ for different waiting time
distributions. In other words, $P(X,t)$ possesses exponentially decaying tails in the
limit of large $|X|/t$ when the distribution of jumps is Laplace distributed.

$P(X,t)$ derived in (\ref{pxt_ldf}) holds for a wide class of waiting time distributions
analytic near zero. This further implies that the rare fluctuations
for a CTRW with Laplace distributed jumps are described by the large deviation
principle \cite{touchette2009large}. In this regard, the case $\beta = 1$ is analogous to the $\beta > 1$ case discussed in Ref.~\cite{barkai2020packets,wang2020large,pacheco2021large} if we restrict our attention solely to the
exponentially decaying fluctuations of $P(X,t)$.
The analogy, however, ends here as for $\beta > 1~ N_0$ increases
linearly with $|X|$ while for $\beta = 1$, the growth is sublinear.
Furthermore, for $\beta > 1$ the PDF exhibits exponentially decaying tails with
logarithmic corrections \cite{barkai2020packets} while for $\beta = 1$ the corrections
are of power-law type. Even though $P(X,t)$ exhibits exponentially decaying tails
for both $\beta > 1$ and $\beta = 1$, different forms of correction term for
the two cases ``hints'' towards a possible transition. Let us now explore the
region $\beta < 1$ to complete our understanding of this transition.

For $\beta < 1$ the distribution of jumps belongs to the class of stretched
exponential distributions \cite{foss2013heavy} which possesses heavy tails as
$\int^\infty_0 dx ~e^{\lambda x} f(x) = \infty ~\forall~\lambda > 0$
and does not admit a large deviation form \cite{touchette2009large}.
It is a well-known result that the family of stretched exponential distributions
satisfies the big jump principle\cite{foss2013heavy,Vezani2019,WangVezanni2019,Vezanni2020,Burioni2020}
\begin{align}
P(x_1+\cdots+x_N \ge X) \stackrel{|X|\to \infty}{\sim}
P(\text{max}\{x_1,...,x_N\} \ge X)
\end{align}
with the right hand side evaluating to $1 - \Big[1-\int^\infty_X dx ~f(x)\Big]^N$
for IID $x_i$. Hence, from (\ref{subor}) we have
\begin{align}
\int^\infty_X dX~P(X,t)
\stackrel{|X|\to \infty}{\sim}
1 -G_t\Big(1-\int^\infty_X dx ~f(x)\Big)
\end{align}
where $G_t(z) = \sum_{N=0}^\infty z^NQ_t(N)$. Furthermore, for large $|X|$ we have
$\int^\infty_X dx ~f(x) \sim 0$, as a result we can analyze $P(X,t)$ in terms of the
behavior of $G_t(z)$ for $z$ in the neighborhood of unity. Now $G_t(1-\eta) \approx
G_t(1) - \fr{\partial G}{\partial z}|_{z=1}\eta$ for $\eta$ small and
$\fr{\partial G}{\partial z} = \sum_{N=1}^\infty N Q_t(N) z^{N-1}$. This implies
$G_t\Big(1-\int^\infty_X dx ~f(x)\Big) \approx 1 - \avg \int^\infty_X dx ~f(x)$
and from here it follows that
\begin{align}
\label{big_jump}
P(X,t) \stackrel{|X|\to \infty}{\sim} \avg f(X).
\end{align}
The above equation implies that the probability of being at a location $X$ at time
$t$ equals the mean number of jumps $\avg$ up to time $t$ times the distribution of a
single jump $f(X)$.
With the distribution of a single jump known,
we only need to estimate the mean number of jumps $\avg$ which in the Laplace domain reads
\cite{klafter2011first} $\mean = \fr{\tl{\psi}_s}{s(1-\tl{\psi}_s)}$, where
$\tl{\psi}_s = \int^\infty_0 dt ~e^{-st} \psi(t)$ is the Laplace transform of $\psi(\tau)$.
For $\psi(\tau)$ analytic near zero (\textit{cf}. (\ref{wt_time})) we have
in the limit $t \to 0$ \cite{suppl}
\begin{align}
\label{avg}
&\avg \approx \fr{C_A\Gamma(A+1)}{\Gamma(A+2)}t^{A+1} + \fr{C_{A+1}\Gamma(A+2)}
{\Gamma(A+3)}t^{A+2}.
\end{align}
The reason to focus on the $t \to 0$ limit is that it
allows us to address the rare fluctuations exhibited by the CTRW at finite times,
that is, $|X|/t \to \infty$. Specifically, in many experiments that show Laplace decay of the PDF, the non-Gaussian behavior was spotted for short enough times~\cite{chechkin2017brownian}. For a long measurement time, the Gaussian center eventually takes over~\cite{pagnini_2022}. 
Notwithstanding
the limited range of validity of (\ref{avg}), $P(X,t)$ derived in
(\ref{big_jump}) holds at arbitrary times,
\begin{figure}
\includegraphics[width=0.45\textwidth]{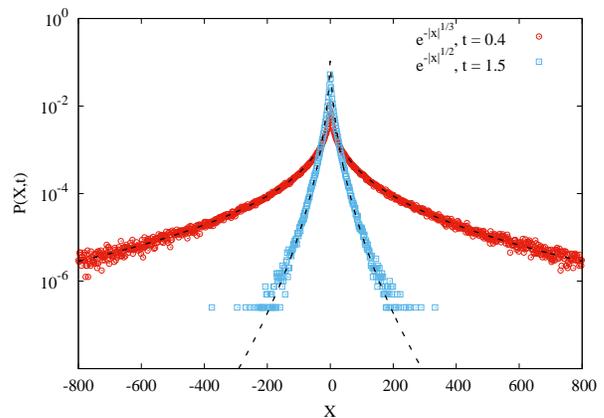}
\caption{Numerically estimated $P(X,t)$ for a CTRW with jumps following a generalized
Gaussian distribution with PDF $f(x) = \fr{\beta}{2\Gamma(1/\beta)}e^{-|x|^\beta}$ with $\beta = 1/3$,
for $\psi(\tau) = \sqrt{2/\pi}e^{-\tau^2/2}$
(red circles) and $\beta = 1/2$ for $\psi(\tau) = \fr{1}{(1+\tau)^2}$ (blue squares).
The black dashed lines represent the analytical form from (\ref{big_jump}).}
\label{fig4}
\end{figure}
and is in excellent agreement with numerical simulations (see Fig.~\ref{fig4}).
It is important to notice that the average number of jumps, $\langle N_t \rangle$, was previously found to be an important quantity determining the tails of the PDF for L\'evy walks~\cite{Vezani2019, Vezanni2020, WangBarkai2020}.

The result in (\ref{big_jump}) shows that the parameter
range $\beta \in (0,1)$ is markedly different from the region $\beta \ge 1$.
In addition, differences in the nature of the PDF of CTRW at $\beta = 1$ and
$\beta > 1$ imply towards the fact that the PDF of a CTRW critically changes at
$\beta = 1$, which is essentially the same value for which the order
parameter $F_\beta(X)$ shows a critical
transition. The evaluation of $P(X,t)$ further corroborates our assertion
of a universal to non-universal transition as seen from the analysis of $F_\beta(X)$
(see Fig.~\ref{fig1}). The fact that $\lim_{|X| \to \infty} F_\beta(X) = 0$ for
$\beta \ge 1$ is analogous to saying that $P(X,t) \sim \exp[-tI(|X|/
t)]$ exists with a nontrivial rate function $I(|X|/t)$ for every
$\beta \ge 1$. 
This rate function $I(z)$ attains a linear growth for large $z$ and, therefore, the universal exponential decay of the PDF, i.e., $P(X,t)\sim e^{-|X|}$.
On the other hand, for $\beta < 1$ we had seen from Fig.~\ref{fig1} that
$\lim_{|X| \to \infty} F_{\beta}(X) = c_\beta >0$. Eq.~\eqref{big_jump} shows that
for $\beta < 1$, the rate function $I(|X|/t)$ is trivially zero. For large $|X|$, the decay of the PDF is stretched exponential, i.e., $P(X,t) \sim
e^{-|X|^\beta}$, that specifically depends on the parameter $\beta$.
Notice that the transition of $\phi_\alpha(t)$ in the long time limit represents the transition from diffusion to subdiffusion and is accompanied by the transition of the PDF (in the $|X|/t\to 0$ limit) from the universal Gaussian form to the $\alpha$-stable L\'{e}vy type that explicitly depends on $\alpha$.

Hopping dynamics which is an intrinsic feature of CTRW, has been ubiquitously observed in polymer melts \cite{schweizer2004theory},
colloidal suspensions \cite{schweizer2003entropic}, rodlike particles through smectic layers 
\cite{lettinga2007self,grelet2008dynamical}, polymer glasses \cite{warren2010deformation}, binary 
mixtures \cite{miotto2021length}, in one, two, and
three spatial dimensions \cite{aoki2013one}, to mention a few.
A characteristic feature of motion in glassy materials 
\cite{chaudhuri2007universal,chaudhuri2008random,wang2009anomalous,wang2012brownian} and at the 
liquid-solid interface \cite{skaug2013intermittent,wang2017three}, where hopping dynamics is observed, has been the exponential decay of the 
tails of the positional PDF. Exponential decay, however, is the rule whenever hopping dynamics is in 
play \cite{barkai2020packets}, making it a universal feature of transport in heterogeneous media. 
But in some situations, like the case of particles with a constant supply of energy (e.g., run and tumble 
particles), the particles can perform really long jumps during their exploration of the heterogeneous media \cite{mori2021first}. 
Our work shows that a critical transition is
expected for any system involving such hops. This critical transition manifests itself at the
level of the positional PDF, where the universality of Laplace tails ceases to exist. While the universal tails are an outcome of an accumulation of many events and the applicability of the large-deviation principle,
the specific tails for $\beta<1$ are determined by one single event, that is, the big-jump principle.

Unlike a diffusion-to-subdiffusion transition which takes place at long times and is accompanied by  
divergences of the mean trapping time, the phase transition reported in the present study is free from such 
divergences. The mean length of a hop can be finite, and the transition is observed at finite times. Furthermore, while the temporal properties of rare
events leading to subdiffusion affect the bulk, rare spatial events manifest themselves mainly in
the tails of the PDF. Interestingly, the temporal features do not affect the spatial dependence of the statistics 
of the rare events, that is, the tails of $P(X,t)$.

\textit{Acknowledgments}: This work was supported by the  Israel Science Foundation Grant No. 2796/20.
RKS thanks the Israel Academy of Sciences and Humanities (IASH)
and the Council of Higher Education (CHE) Fellowship.

\bibliography{ctrw.bib}

\end{document}


\title{Supplementary Material for Universal to Non-Universal Transition of the statistics of
Rare Events During the Spread of Random Walks}

\author{R. K. Singh}
\email{rksinghmp@gmail.com }
\affiliation{Department of Physics, Bar-Ilan University, Ramat-Gan 5290002, Israel}

\author{Stanislav Burov}
\email{stasbur@gmail.com }
\affiliation{Department of Physics, Bar-Ilan University, Ramat-Gan 5290002, Israel}


\maketitle

\newcommand{\fr}{\frac}
\newcommand{\tl}{\tilde}
\newcommand{\avg}{\langle N_t \rangle}
\newcommand{\mean}{\langle \tilde{N}_s \rangle}

\section{Analogy between time and space}
While studying the diffusion to subdiffusion transition for continuous time random walks (CTRWs) with
waiting times following $\psi(\tau) \sim \tau^{-1-\alpha}$ ($\tau\to\infty$) we have used the
quantity
\begin{align}
\phi_\alpha(t) = \Big\langle \fr{\text{max}\{\tau_1, ..., \tau_{N_t}\}}{\sum^{N_t}_{i=1} \tau_i}\Big\rangle
\label{phi_t}
\end{align}
and analyzed its long-time behavior. While considering the maxima of
waiting times $\tau_i$, bounded above by $t$, we do not explicitly
account for the backward reference time $B_t = t-\sum^{N_t}_{i=1}\tau_i$
\cite{holl2020extreme} for a couple of reasons. First is that as $\phi_\alpha(t)$
is a function
of $t$, occurrence of even a large $B_t$ (expected for power
law waiting times) would not change its value because
$\tau_i$ are sampled up to the last jump. Secondly, exclusion
of $B_t$ allows us to get away with the correlations which
will appear due to the fact that the time of measurement
is fixed at $t$ \cite{holl2020extreme}. This has the further advantage of putting time and space
on equal footing as the quantity to study spatial fluctuations
\begin{align}
        F_\beta(X) = \Big\langle\fr{\text{max}\{|x_1|,...,|x_{N_t}|\}}
        {\sum^{N_t}_{i=1}|x_i|}\Big\rangle.
\end{align}
is analogous to $\phi_\alpha(t)$ with $\tau_i$ replaced by $|x_i|$. In terms of fluctuations, we
focus on the absolute value of location of the CTRW upto the last jump in $F_\beta(X)$. Similarly,
$\phi_\alpha(t)$ takes into account the waiting times only upto the last jump taking place before
the observation time $t$. The analogy of $\phi_\alpha(t)$ and $F_\beta(X)$ sets the premise to
study phase transitions at finite times once we start looking at the rare fluctuations of a
CTRW.

\section{Mean number of jumps near $t=0$}
\begin{figure}
\includegraphics[width=0.45\textwidth]{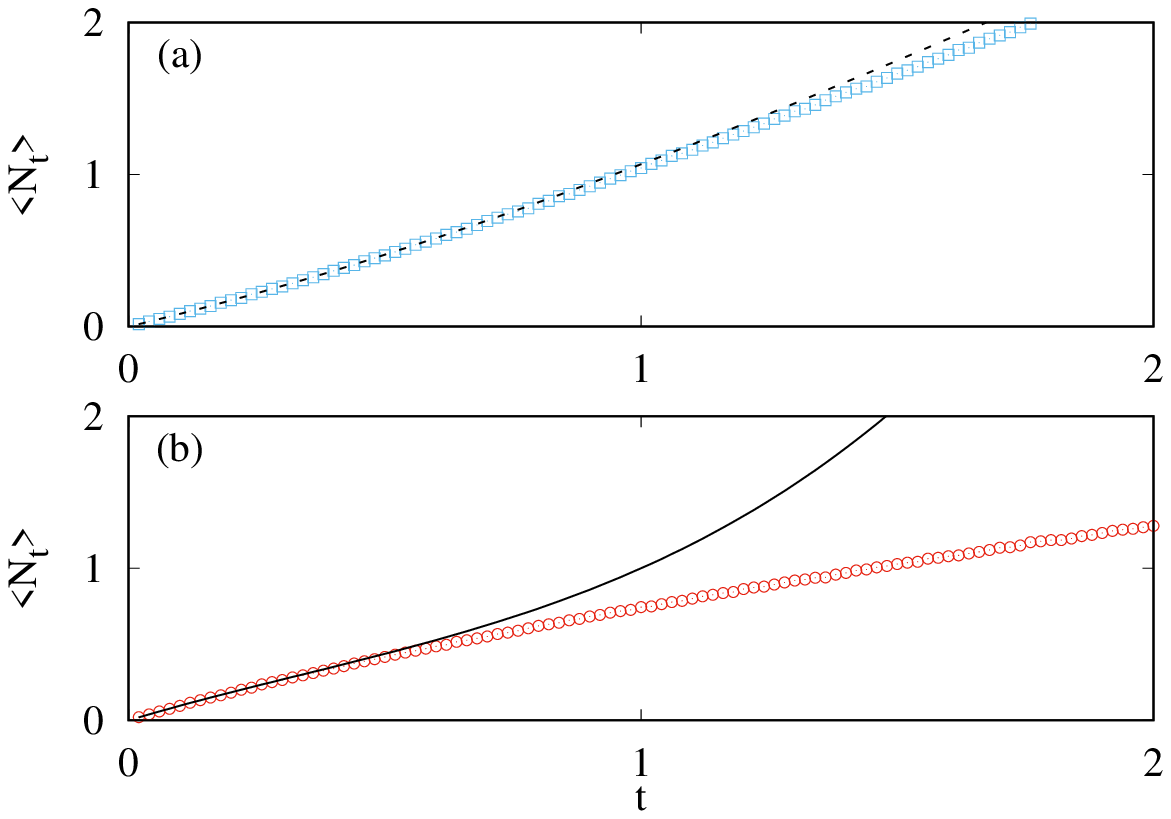}
\caption{Comparison of numerically estimated $\avg$ against the solution
in (\ref{avg}) for (a) half Gaussian distribution: $\psi(\tau) = \sqrt{\fr{2}{\pi}}
e^{-\tau^2/2}$, and (b) power law distribution $\psi(\tau) = \fr{1}{(1+\tau)^2}$.
The symbols are numerical calculations and lines are $\avg$ evaluated
from (\ref{avg}).}
\label{fig3}
\end{figure}
~\\
The mean number of jumps in Laplace space is
$\mean = \fr{\tl{\psi}_s}{s(1-\tl{\psi}_s)}$ \cite{klafter2011first}
and for waiting time distribution analytic near zero $\psi(\tau)
\stackrel{\tau \to 0}{\sim} C_A \tau^A + C_{A+1} \tau^{A+1} +
C_{A+2} \tau^{A+2} + \cdots$ we have in time domain
\begin{align}
\label{avg}
&\avg \approx \fr{C_A\Gamma(A+1)}{\Gamma(A+2)}t^{A+1} + \fr{C_{A+1}\Gamma(A+2)}
{\Gamma(A+3)}t^{A+2} + \fr{C_{A+2}\Gamma(A+3)}{\Gamma(A+4)}t^{A+3}
+ \fr{C_{A}^2\Gamma^2(A+1)}{\Gamma(2A+3)}t^{2A+2} \nonumber\\
&+ \fr{2C_AC_{A+1}\Gamma(A+1)\Gamma(A+2)}
{\Gamma(2A+4)}t^{2A+3} + \fr{C^3_{A}\Gamma^3(A+1)}{\Gamma(3A+4)}t^{3A+3}.
\end{align}

We compare the approximate value of $\avg$ evaluated from (\ref{avg}) against
numerical calculations in Fig.~\ref{fig3} and find that the approximate form
in (\ref{avg}) captures the true behavior only at small times. The domain of
the validity, however, depends on the exact nature of the distribution. For example,
when the distribution of waiting times is half Gaussian, that is, $\psi(\tau)
= \sqrt{\fr{2}{\pi}}e^{-\tau^2/2}$, we have $A = 0, C_A = \sqrt{\fr{2}{\pi}},
C_{A+1} = 0, C_{A+2} = -\sqrt{\fr{1}{2\pi}}$ and it is evident from Fig.~
\ref{fig3} (a) that the approximate form derived in (\ref{avg}) agrees with
numerically estimated $\avg$ upto $t \approx 1$. On the other hand, for
the power law distribution $\psi(\tau) = \fr{1}{(1+\tau)^2}$ we find that the usefulness of (\ref{avg}) is reduced to half the range, that is, $t \in (0,1/2)$.
The reason for this difference  is that the small time behavior of $\psi(\tau)$
does not capture jumps taking place at finite times.

\bibliography{suppl.bib}